\documentclass[10pt,letterpaper,twocolumn,superscriptaddress]{article} 
\usepackage{ol2}
\usepackage[draft]{hyperref}
\usepackage{amsmath}
\usepackage[section]{placeins}
\usepackage{amsmath}
\usepackage{amssymb}
\usepackage{amsfonts}
\usepackage{dsfont}
\usepackage{graphicx}
\usepackage{times,txfonts}
\newcommand{\braket}[2]{\langle#1|#2\rangle} 
\usepackage{soul}
\definecolor{lightblue}{RGB}{185,210,248}
\sethlcolor{lightblue}

\begin{document}

\twocolumn[ 

\title{Limitations to the determination of a Laguerre-Gauss spectrum via projective, phase-flattening measurement}

\author{Hammam Qassim,$^{1}$ Filippo M. Miatto,$^{1}$ Juan P. Torres,$^{2}$ Miles J. Padgett,$^{3}$  Ebrahim Karimi,$^{1,*}$ and Robert W. Boyd$^{1,3,4}$}

\address{ $^1$Department of Physics, University of Ottawa, 150 Louis Pasteur, Ottawa, Ontario, K1N 6N5 Canada\\
$^2$ ICFO-Institut de Ciencies Fotoniques, 08860 Castelldefels (Barcelona), Spain\\
$^3$ School of Physics and Astronomy, SUPA, University of Glasgow, Glasgow G12 8QQ, United Kingdom\\
$^{4}$ Institute of Optics, University of Rochester, Rochester, New York, 14627, USA\\
$^*$Corresponding author: ekarimi@uottawa.ca}

\begin{abstract}
One of the most widely used techniques for measuring the orbital angular momentum components of a light beam is to flatten the spiral phase front of a mode, in order to couple it to a single-mode optical fiber. This method, however, suffers from an efficiency that depends on the orbital angular momentum of the initial mode and on the presence of higher order radial modes. The reason is that once the phase has been flattened, the field retains its ringed intensity pattern and is therefore a nontrivial superposition of purely radial modes, of which only the fundamental one couples to a single mode optical fiber. In this paper, we study the efficiency of this technique both theoretically and experimentally. We find that even for low values of the OAM, a large amount of light can fall outside the fundamental mode of the fiber, and we quantify the losses as functions of the waist of the coupling beam of the orbital angular momentum and radial indices. Our results can be used as a tool to remove the efficiency bias where fair-sampling loopholes are not a concern. However, we hope that our study will encourage the development of better detection methods of the orbital angular momentum content of a beam of light.\newline
\end{abstract}
\ocis{070.2580, 230.6120,  070.6042. }

 ] 
\section{Introduction}
\noindent Structured light beams have wide applications in technologies such as lithography, nanoscopy, spectroscopy, optical tweezers and quantum cryptography~\cite{grosjean:07,hell:07,he:95,paterson:01,molina:07}. Among these, beams with helical phase fronts $\exp{(i\ell\phi)}$, where $\ell$ is an integer number and $\phi$ is the azimuthal angle in polar coordinates, are of particular interest since they can be used for classical~\cite{gibson:04,wang:12,siddharth:13} and quantum communications~\cite{molina:07}. These beams carry a well-defined value of optical orbital angular momentum (OAM) $\ell\hbar$ per photon along the propagation direction. Due to these proposed applications, there are fervent attempts to design innovative devices to generate  such beams. Until now, possible solutions include spiral phase plates~\cite{beijersbergen:94}, computer-generated holograms imprinted onto spatial light modulators (holographic approach)~\cite{bazhenov:92,ngcobo:13}, mode converters (cylindrical lenses)~\cite{allen:92}, $q$-plates (nonuniform liquid crystal plates)~\cite{marrucci:06,karimi:09}, and some types of OAM-sorters~\cite{berkhout:10,mohammad:13}. These solutions are practical and widely used in various experimental realizations, and are implemented both in classical and quantum regimes. However, with the exception of mode converter and a hologram with an intensity mask~\cite{leach:05,eliot:13}, the above methods do not generate a pure Laguerre-Gauss mode~\cite{karimi:07,dennis:09}. In some cases, the reverse process can be used to \emph{detect} the spectrum of OAM of an unknown beam, where each mode is coupled to a single mode optical fiber (SMOF) after its azimuthal phase dependence has been flattened. Such a method, was first introduced by Mair et al.~\cite{mair:01} in the quantum domain and then used commonly in the classical regime. This technique might sound accurate, but as we will show, its shortcoming is that the OAM bandwidth that can be measured has a bias that depends on the characteristics of the beam. As a consequence, the bias for an unknown beam cannot be removed. Moreover, the detection efficiency for high OAM modes can be extremely low, making it seem like those components are very weak. This issue is particularly important for those experiments that rely on a high detection efficiency, for example, experiments that aim at maximizing the heralding efficiency, or at closing a detection loophole~\cite{brunner:10,dada:11}, or at characterizing a state by measuring each of its OAM components separately~\cite{torres:03,pires:10}. In this letter, we study projective measurements based on phase-flattening followed by coupling into a SMOF. We examine our theoretical model experimentally for various mode projections, and we verify the trends in coupling efficiencies.\newline

\section{Theoretical analysis}
In our analysis we use Laguerre-Gauss (LG) modes, which are characterized by two indices: the radial index $p$ (nonnegative integer) and the azimuthal number $\ell$ (integer), which are associated to the number of radial nodes and to the OAM value, respectively. The LG modes are a complete and orthonormal family of solutions of the paraxial wave equation, i.e. (in Dirac notation) $\braket{p',\ell'}{p,\ell}=\delta_{p',p}\,\delta_{\ell',\ell}$, and in the position representation at the pupil they are given by
\begin{align}
	\label{eq:lgs}
	\mbox{LG}_{p,\ell}(r,\phi)&:=\sqrt{\frac{ 2^{|\ell|+1}p!}{\pi w_0^2\,(p+|\ell|)!}}\,\left(\frac{r}{w_0}\right)^{|\ell|} e^{-\frac{r^2}{w_0^2}} L_{p}^{|\ell|}\left(\frac{2r^2}{w_0^2}\right)\,e^{-i\ell\phi},
\end{align}
where $r, \phi$ are the transverse cylindrical coordinates, $w_0$ is the beam waist radius at the pupil and $L_{p}^{\ell}(.)$ is the generalized Laguerre polynomial. The devices listed above can generate LG modes with limited fidelity. The most convenient and commonly used method is the holographic approach, with an embedded intensity masking. However, a \emph{mode-cleaning filter cavity} can be used to increase fidelity of the generated mode~\cite{granata:10,carbone:13}. 

\subsection{Projecting on LG modes}
To perform a projective measurement, the mode $\mbox{LG}_{p,\ell}$ (in our case generated by an SLM) is imaged onto a \emph{different} conjugate mode, $\mbox{LG}_{p',\ell'}^{*}$, and the resulting field is propagated and coupled into a SMOF in the far-field, which selects only the near Gaussian component. Imaging onto an SLM is described by taking the product of the two modes, i.e.  $\mbox{LG}_{p,\ell}(\mathbf r_\bot)\,\mbox{LG}_{p',\ell'}^{*} (\mathbf r_\bot)$ where $\mathbf r_\bot$ stands for the transverse coordinates. The far-field distribution becomes a polynomial-Gaussian function given by a 2D-Fourier transform:
\begin{align}
	\label{eq:2dFF}
{\cal F}_{p,\ell}\left(\rho,\varphi\right)={\cal{FT}}\left[\mbox{LG}_{p,\ell}(\mathbf r_\bot) {\mbox{LG}_{p',\ell'}^*(\mathbf r_\bot)}\right],
\end{align}
where ${\cal{FT}}$ stands for the 2D-Fourier transform, and $\rho$ and $\varphi$ are the cylindrical coordinates in the far field.
\begin{figure}[!t]
	\centering
	\includegraphics[width=\columnwidth]{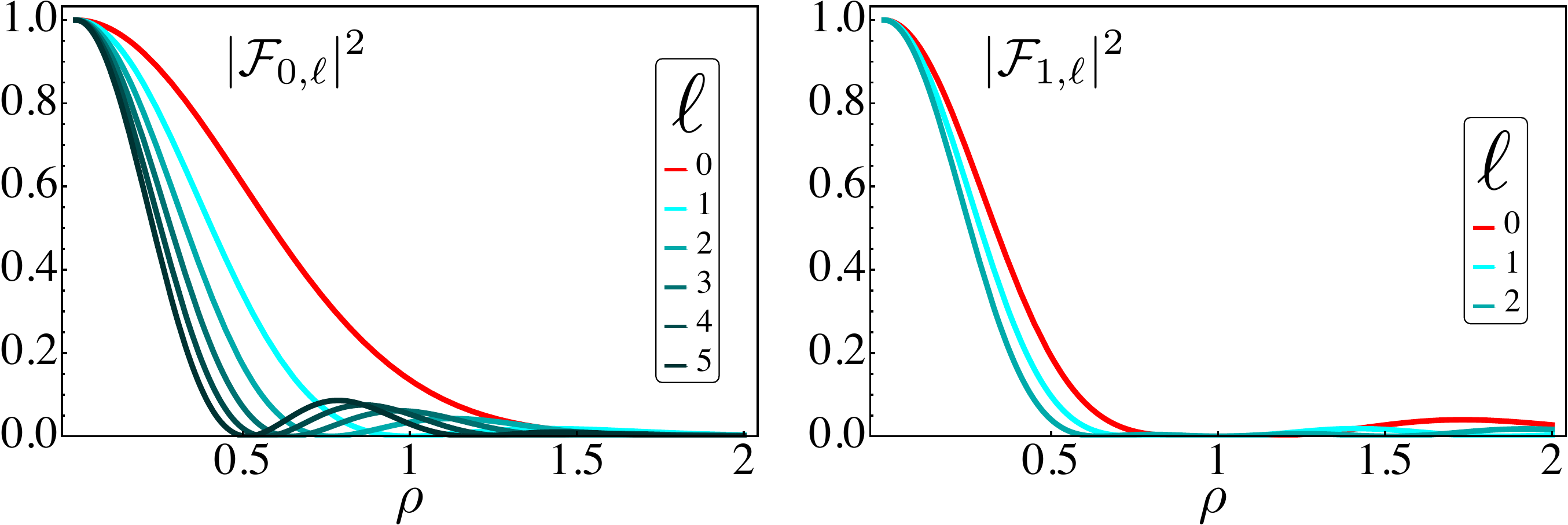}
	\caption{(Color online) Intensity of the $p=0$ (left) and $p=1$ (right) modes at the input to the fiber. As an effect of diffraction, local maxima at the periphery gain intensity as $p$ and $|\ell|$ increase. Here $\rho$ is in units of $a_0=(\sqrt{2}\lambda f)/(\pi w_0)$, which is the natural scaling factor in the far-field of the lens.}
	\label{fig:fig1}
\end{figure}
The fact that a SMOF only supports the $\mathrm{TEM}_{00}$ mode limits this technique to the case in which $\ell'=\ell$. Moreover, as oscillating radial phases would alter the coupling to the SMOF, we also choose $p'=p$. Due to the absence of any angular dependence after the phase flattening stage, the 2D-Fourier transform ${\cal FT}$ can be simplified into the Hankel transform of order zero, i.e.
 \begin{align}
	\mathcal F_{p,\ell}(\rho,\varphi)=\frac{2\pi e^{\frac{i\pi}{\lambda f}\rho^2}}{i\lambda f}\int_0^\infty rdr|\mathrm{LG}_{p,\ell}(\mathbf{r}_\bot)|^2J_0\left(\frac{2\pi}{\lambda f}r\rho\right).
\end{align}
In Fig.~\ref{fig:fig1} we show some examples of transverse intensity at the fiber for several values of $p$ and $\ell$. Notice that the beams have a Gaussian-like shape with local maxima at the periphery, which give rise to a ringed pattern in the transverse plane. As $|\ell|$ and $p$ become larger, the beam intensity distribution moves to the outer rings. This is related to the effect that a larger phase-flattened doughnut beam is turned into a smaller and weaker central spot at the far field, which has been studied and discussed for special cases in \cite{slussarenko:09,gabriel:07}.

\begin{figure}[!t]
	\centering
	\includegraphics[width=\columnwidth]{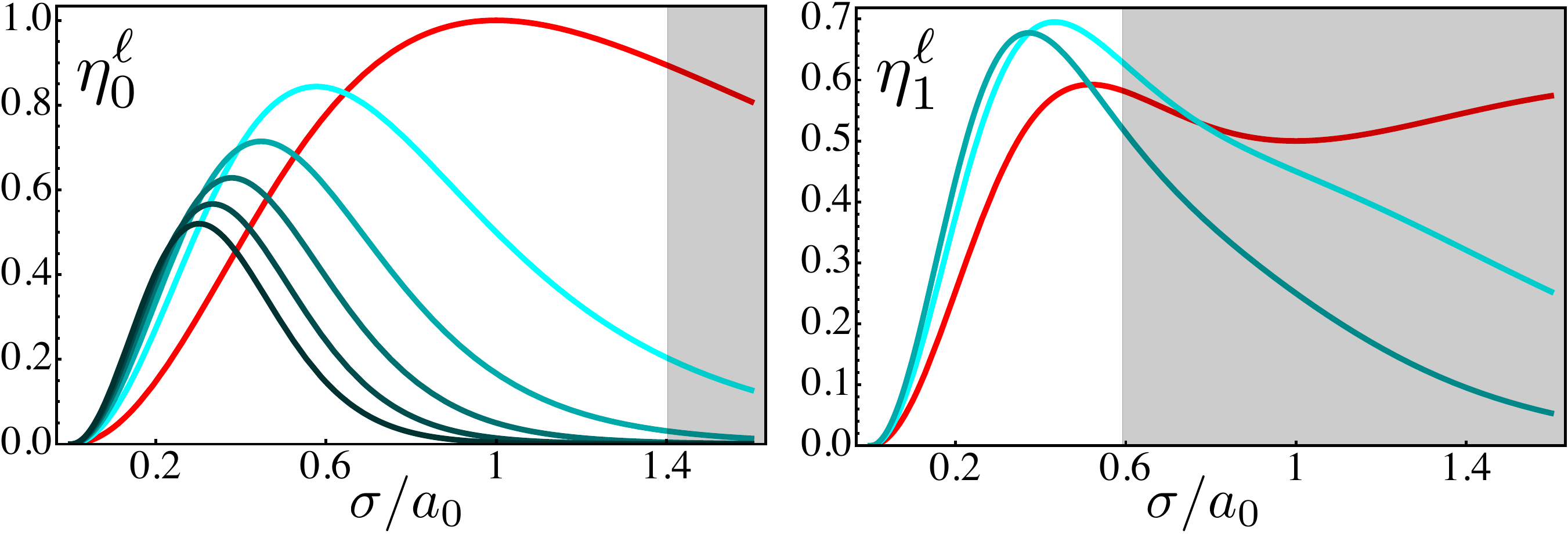}
	\caption{(Color online) Coupling efficiency for projective measurements for the modes in Fig.~\ref{fig:fig1}. For a given choice of optics, the coupling efficiency shows a bias dependent on the order of the transverse modes. The horizontal axis is scanned by changing $w_0$, as $a_0$ is inversely proportional to $w_0$. The shaded box indicates the region limited by the active area of the SLMs (see experimental section).}
	\label{fig:fig2}
\end{figure}

The coupling efficiency to a SMOF, then, is given by the overlap of the Gaussian mode supported by the fiber and the far-field distribution calculated in \eqref{eq:2dFF}:
\begin{align}\label{eq:projection}
	\eta^{\ell}=\frac{2}{\pi \sigma^2}\left|\int_{0}^{\infty}\rho\,d\rho\int_{0}^{2\pi}d\varphi\,\,{\cal F}_{\ell}\left(\rho,\varphi\right)\,e^{-\frac{\rho^2}{\sigma^2}}\right|^2,
\end{align}
where $\sigma$ is the beam waist radius of the SMOF Gaussian mode. It is worth mentioning that the mode of a SMOF can be approximated with a Gaussian beam. Here we give the results for $p=0$ and $p=1$:
\begin{align}
	\eta_0^\ell&=\frac{|\ell|!^2}{(2|\ell|)!}A^{2|\ell|+1}B\\
	\eta_1^\ell&=\frac{(|\ell|+1)!|\ell|!}{4(2+3|\ell|)(2|\ell|)!}A^{2|\ell|+1}B(A^2+B^2(|\ell|+1))^2,
\end{align}
where $A=2/(1+\frac{\sigma^2}{a_0^2})$ and $B=2/(1+\frac{a_0^2}{\sigma^2})$, and $a_0=(\sqrt{2}\lambda f)/(\pi w_0)$ is the natural scaling factor at the fiber. Notice that it is only the ratio $\sigma/a_0$ that matters, as it should be. These results are shown in Fig.~\ref{fig:fig2}, where it is possible to see that the highest coupling efficiency for different modes is achieved for different values of the waist at the fiber, which can be tuned by adjusting the focal length of the Fourier lens.\newline

\subsection{Projecting on spiral modes}
An alternate and less desirable solution that we explore only theoretically is to project onto a purely spiral field $e^{i\ell\varphi}$ (which can be implemented with a pitchfork hologram on an SLM), whereby the effect is to simply cancel out the spiral phase from an initial LG mode, in which case the field at the fiber is given by
$
{\cal F}_{p,\ell}\left(\rho,\varphi\right)={\cal{FT}}\left[\mbox{LG}_{p,\ell}(\mathbf r_\bot) e^{i\ell\phi}\right].
$
This equation can be solved analytically, recall that the Eq.~(\ref{eq:2dFF}) has an analytical solution only once a value of $p$ is specified. However, for this specific case the coupling efficiency is given by
\begin{align}\label{eq:projectionflattening}
	\eta_{p}^{\ell}=\left|{\mathcal{N}_p^{\ell}\sum_{j=0}^{p}(-1)^{j}{p\choose j}a_j^{\ell} f_j^{p,\ell}}\right|^2
\end{align}
with
\begin{align}\label{eq:projectiondefinition}\nonumber
	f_j^{p,\ell}=&\prod_{k=1}^p\bigl(|\ell|+k+k\,H(j-k)\bigr)\cr
	a_j^{\ell}=&\frac{\sqrt{2\pi}\,\sigma/a_0}{\left(1+(\sigma/a_0)^2\right)^{\frac{|\ell|}{2}+j+1}},
\end{align}
where $\mathcal{N}_p^{\ell}=\sqrt{\frac{2^{|\ell|+1}}{\pi (p+|\ell|)!\,p!}}\,\Gamma\left(\frac{|\ell|}{2}+1\right)$ is the normalization function. $H(j-k)$ in $f_j^{p,\ell}$ is the unit step function: its value is 0 for $j<k$ and $1$ for $j\geq k$, and $\Gamma$ is the gamma function, respectively. As was expected, the coupling efficiency $\eta_{p}^{\ell}$ depends on the ratio between the beam waist radius of the SMOF $\sigma$ and the size of the field at the fiber position $a_0$.\newline
\begin{figure}
	\centering
	\includegraphics[width=\columnwidth]{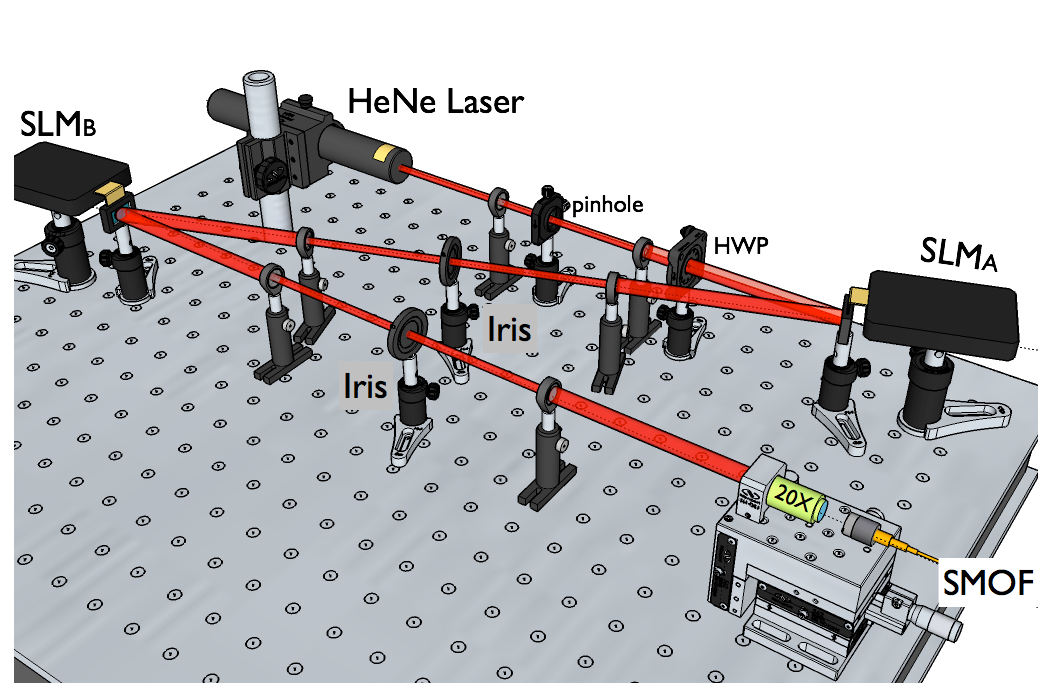}
	\caption{(Color online) Experimental setup for generating and detecting photon transverse states. A linearly polarized HeNe laser beam is spatially cleaned with two lenses and a pinhole. A half-wave plate (HWP) optimizes the first order of diffraction on SLM$_\text{A}$, since SLMs are polarization dependent. The mode $\mbox{LG}_{p,\ell}(\mathbf{r}_\bot)$ produced by SLM$_\text{A}$ is then projected on the mode $\mbox{LG}_{p,\ell}(\mathbf{r}_\bot)^{*}$ on SLM$_\text{B}$. The resulting far field is coupled into a single mode optical fiber (SMOF). We implement two $4f$-system with unit magnification and a microscope objective to image SLM$_\text{A}$ on SLM$_\text{B}$ and SLM$_\text{B}$ on the microscope objective. Irises are used to select the first order of diffraction at the far-field plane of SLMs, where higher order of diffraction are well separated.}
	\label{fig:fig3}
\end{figure}

\section{Experimental results}
In order to verify the above theory, we prepared an experimental setup (Fig.~\ref{fig:fig3}) in which we examined the projective measurement method for different sets of transverse modes with varying beam sizes.  A linearly polarized light beam of a HeNe laser is spatially cleaned, and illuminates the first of two Pluto HOLOEYE SLMs (SLM$_A$), to generate the initial $\mbox{LG}_{p,\ell}(\mathbf{r}_\bot)$ mode. This is then imaged on a second SLM (SLM$_B$) via a $4f$-system with unit magnification, where the mode is projected onto $\mbox{LG}_{p,\ell}(\mathbf{r}_\bot)^*$. We used intensity masking to encode transverse modes with high fidelity~\cite{eliot:13,vincenzo:13}. The product field is finally coupled to a SMOF with mode diameter of $\simeq4.8\,\mu$m and a numerical aperture $\mathrm{NA}=0.12$ at the far-field of a $20\times$ microscope objective ($f=9$ mm and $\mathrm{NA}=0.40$). In order to normalize the coupling efficiency for different modes, we used a Newport power meter with two read out heads to record both the coupling efficiency and the power of the field just before the fiber simultaneously. Recall that due to the intensity masking different modes have different generation and detection efficiencies, for more details see Ref.~\cite{eliot:13,flamm:13}. An automatic program optimized the center of the holograms on the both SLMs and the coupling efficiency with the SMOF. The pixel size and active area of the SLMs were $8\,\mu$m and $15.36\,\mbox{mm}\times8.64\,\mbox{mm}$. These characteristics set the limits of the range of beam waists and mode numbers that could be investigated.

\begin{figure}[!t]
	\includegraphics[width=\columnwidth]{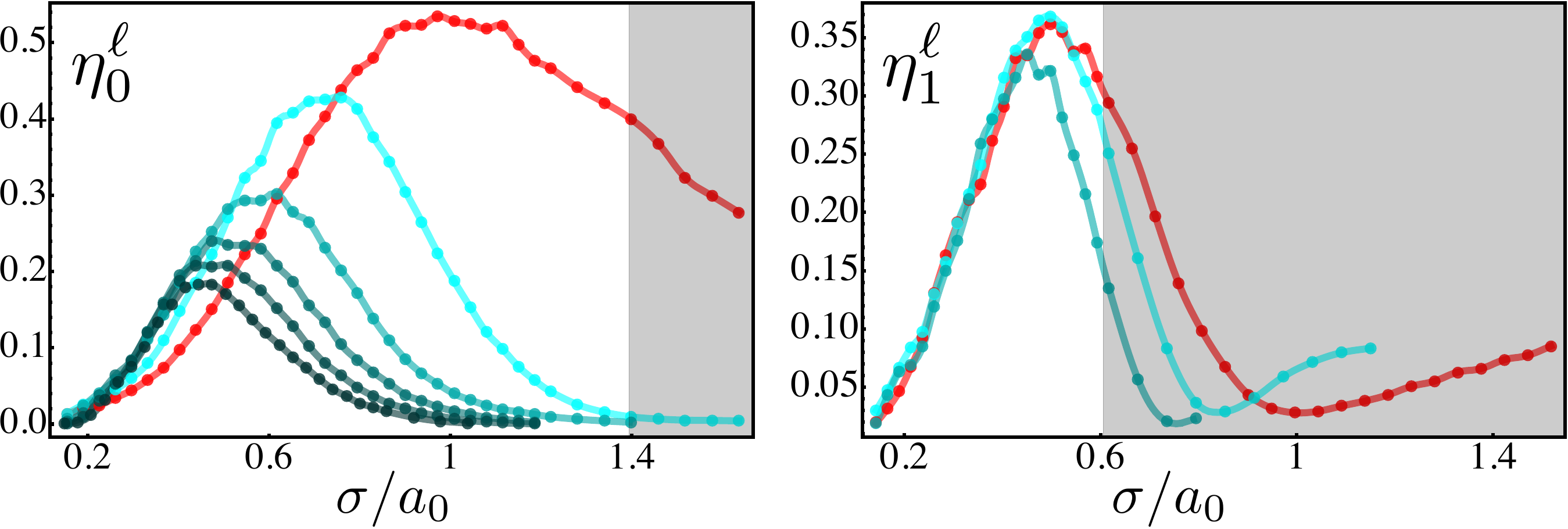}
	\caption{(Color online) Experimentally measured overall coupling efficiency for the modes shown in Fig.~\ref{fig:fig1}: (left) $\ell=0\dots5$ and (right) $\ell=0\dots2$. The shaded regions indicate a domain in which the effective beam size exceeds the active area of the SLM, resulting in unreliable data.}
	\label{fig:fig4}
\end{figure}

Figure~\ref{fig:fig4} shows the experimental results, to be compared with the coupling efficiency shown in Fig.~\ref{fig:fig2}. Aside from an overall multiplicative efficiency of about $50\%$ (which comprises reflection and scattering by microscope objective and fiber), the observed data (Fig.~\ref{fig:fig4}) and the theoretical model (Fig.~\ref{fig:fig2}) agree, especially in those regions where the SLMs resolution and active area do not affect the quality of the generated and projected beams. The region below 0.2 $\sigma/a_0$ is limited by resolution, as too few pixels are used. On the opposite end of the horizontal axis the beams eventually fall out of the the active area. These regions are indicated by a shaded area in the figures. Due to truncation mainly induced by the microscope objective, there a small deviation for the case of $p=1$ at large beam waist size with respect to the theoretical calculation. However, the spread of these curves is an indication that the coupling efficiency differs for different initial modes and that therefore the spectrum that is ultimately measured is likely not representative of the true OAM distribution of the beam. We can deduce that projective measurement methods should be used with care. A possible solution could be to calibrate the coupling efficiency for different transverse modes and then post-process the measurement data, but even in this case, if the radial distribution of the initial field is unknown, the bias may not be removable, as the radial decomposition depends on the waist that is chosen for the modes. Of course it is also true that a linear superposition of LG beams leads to an inaccurate result, since the projective measurement gives a bias among projection of different pure OAM states. \newline

\section{Conclusions}
In conclusion, we studied the efficiency of projective measurement as a method to characterize the transverse mode of a light beam. Our analysis can be summarized in two important messages. The first is that although the coupling efficiency is modal- and beam waist-dependent, the bias that is induced might be removed in post-processing after a careful calibration. Of course, issues may arise in the context of an experiment aimed at violating Bell's inequalities: post-processed results could be regarded as an artificial manipulation of the data, and detection-related loopholes might be called into consideration. The second message builds on the fact that the radial modal content of the initial beam depends on the waist $w_0$ that is chosen for the decomposition, and the optimal choice (i.e. the one that results in the least number of radial modes) for a general beam could be found only after further measurements. However, as one then should have to calibrate for this optimal size, this is clearly not an ideal procedure, especially in the context of quantum optics, where there might be a scarce number of photons available). It is our hope that this work will motivate the search of new and better measurement techniques for OAM and more generally for the transverse radial modes of a light beam.\newline
 
\section{Acknowledgments}
The authors thank Prof. Gerd Leuchs for fruitful discussions. H.Q., F.M.M., E.K. and R.W.B. acknowledge the support of the Canada Excellence Research Chairs (CERC) Program. M.J.P. and R.W.B. were supported by the DARPA InPho program. M.J.P. supported by the royal society.

\end{document}